\documentclass[12pt]{article}
\usepackage{a4wide}
\usepackage{amsmath,amsfonts,amssymb,amsthm,graphicx}

\newtheorem{theorem}{Theorem}[section]
\newtheorem{remark}[theorem]{Remark}

\newtheorem{lemma}[theorem]{Lemma}

\theoremstyle{definition}

\numberwithin{equation}{section}

\usepackage{color}
\newcommand{\blue}{\textcolor{blue}}

\def\QED{\mbox{\rule[-1.5pt]{6pt}{10pt}}}

\definecolor{violet}{rgb}{0.219,0.112,0.37}

\begin{document}

\vskip 0.5cm

\noindent \phantom{.} {\qquad \hfill\blue{{\textbf{Submitted to JMAG}}}}

\vskip 1.0cm

\begin{center}

{\Large{\textbf{Schr\"{o}dinger operator in the limit of shrinking wave-guide cross-section and
singularly scaled twisting}}}

\vspace{0.5cm}

\textbf{C\'{e}line Gianesello} \footnote{E-mail: Celine.Gianesello$@$cpt.univ-mrs.fr}

\vspace{0.2cm}

{Universit\'e du Sud, Toulon-Var et\\ Centre de Physique
Th\'eorique - UMR 6207 \\ Luminy - Case 907,
13288 Marseille, Cedex 09, France}

\vspace{0.5cm}

\textbf{Abstract}
\end{center}

Motivated by the method of self-similar variables
for the study of the large time behaviour of the heat equation in twisted wave-guides,
we consider a harmonic-oscillator-type operator in hard-wall three-dimensional wave-guides whose non-circular cross-section
and the support of twisting  diminishing \textit{simultaneously} to zero.

Since in this limit the strength of the twisting increases to infinity and its support shrinks to the point,
we show that the corresponding Schr\"{o}dinger operator converges in a suitable norm-resolvent
sense to a one-dimensional harmonic-oscillator operator on the reference line of the wave-guide,
subject to some extra Dirichlet boundary condition at the twisting point support.

\tableofcontents

\newpage
\section{Introduction}
\begin{figure}[h]
\begin{center}
\includegraphics[scale=0.22]{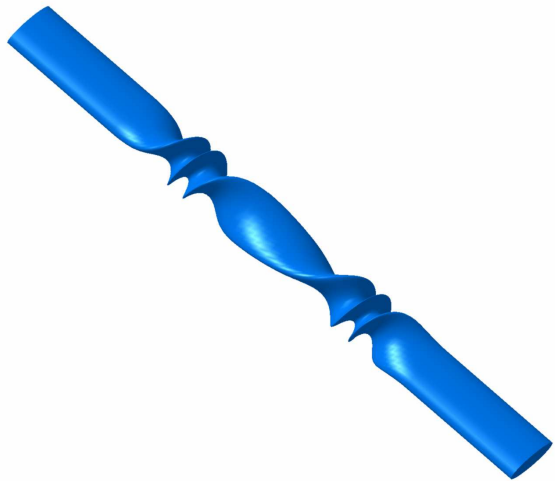}
   \caption{A twisted waveguide with non circular cross section}\label{Fig1}
\end{center}
\end{figure}

\noindent
While the effect of bending in quantum wave-guides
has been studying since a long time,
see e.g. \cite{DuEx}, \cite{ExSh}, \cite{CDFK},
that of twisting has been observed only recently \cite{EKK}.
It is well known that the curvature of the reference curve
leads to some kind of attractive interaction, which gives rise to geometrically induced bound states.
On the other hand, the recent results show that local non-trivial rotations~$\theta$ of the wave-guide with
non-circular cross-section (\textit{twisting}, see Figure~\ref{Fig1}) generate Hardy-type estimates for energy spectrum,
which in particular exclude the existence of bound states. Therefore, one deals with an interesting spectral-geometric
interplay in simultaneously bent and twisted tubes -- see \cite{Kr08} for a review and references.
%

Another important consequence of the Hardy-type inequalities
has been studied recently in \cite{KrZu10}
in the context of the heat equation.
The authors show that the heat semigroup acquires
an extra decay rate due to twisting,
as compared to the straight (untwisted) wave-guide.
The robustness of this effect of twisting has been subsequently
demonstrated on other related models,
such as planar wave-guides with twisted boundary conditions
\cite{KK2}, \cite{KrZu11},
and strip-like negatively curved manifolds
\cite{K3}, \cite{KK3}.

The approach of \cite{KrZu10} is based on the method
of self-similar variables and weighted Sobolev spaces,
which reduce the problem of large-time behaviour
of solutions to the heat equation
to the study of the convergence of the family
of singularly scaled Schr\"odinger-type operators
\begin{equation}\label{self-Hamiltonian}
H_{\varepsilon} =
-(\partial_1-\sigma_\varepsilon\,\partial_\tau)^2
+\frac{x^2_1}{16}
-\frac{1}{\varepsilon^2} \, \Delta_D^\omega
- \frac{1}{\varepsilon^2} \, E_1
\quad \mbox{in}\quad L^2{(\Omega_0)} ,
\end{equation}
subject to Dirichlet boundary conditions,
as the singular parameter~$\varepsilon$
(playing the role of inverse exponential of the self-similar time)
tends to zero.
Here $\Omega_0:=\mathbb{R}\times \omega$ is a straight tube
(to which the twisted wave-guide can be mapped by using suitable
curvilinear coordinates) of cross-section $\omega\subset \mathbb{R}^2$,
$-\Delta_D^\omega$ and $\partial_\tau$ denote the Laplacian
and angular derivative in~$\omega$, respectively,
$E_1$ is the first eigenvalue of the Dirichlet Laplacian in $L^2(\omega)$
and $\sigma_\varepsilon$ is the singularly scaled twisting:
\begin{equation}\label{self-twist}
  \sigma_\varepsilon(x)
  : = \frac{1}{\varepsilon} \ \dot{\theta}\left(
  \frac{x_1}{\varepsilon}
  \right)
  .
\end{equation}
Note that the appearance of~$\varepsilon$ in \eqref{self-Hamiltonian}
is such as if the tube were shrinking to the reference line
as $\varepsilon \to 0$,
while the velocity of the twisting angle~$\theta$ in~\eqref{self-twist}
grows and its support diminishes in the limit.
The overall feature of \eqref{self-Hamiltonian}
is therefore very singular in the limit $\varepsilon \to 0$.

As the main ingredient in the proof of the extra decay rate
of the heat semigroup, the authors of \cite{KrZu10} prove
a strong-resolvent convergence of~$H_{\varepsilon}$ as $\varepsilon \to 0$
to the one-dimensional harmonic-oscillator operator
\begin{equation}\label{h0.intro}
h_D =-\frac{d^2}{dx_1^2}+\frac{x_1^2}{16} \ \
\quad \mbox{in}\quad L^2(\mathbb{R}) \, ,
\end{equation}
subject to a supplementary Dirichlet condition at $x_1=0$
if and only if the tube is non-trivially twisted.
It is in fact this decoupling condition which is responsible
for the faster decay rate of solutions to the heat equation
in twisted tubes, since the lowest eigenvalue of \eqref{h0.intro}
determines the decay rate and the former is three times
greater if the supplementary Dirichlet condition is present.

In this paper we show that the convergence of~$H_{\varepsilon}$
to $h_D$ as $\varepsilon \to 0$ actually holds
in a (suitable)\textit{ norm-resolvent} sense
(taking into account the fact that the operators
act on different Hilbert spaces).
Our approach (different from that of \cite{KrZu10})
essentially uses the technique of \cite{BGW} and,
apart from giving the operator convergence in a better topology,
it enables us to establish the rate of convergence.
We also note that the question of the validity of
the norm-resolvent convergence was explicitly raised in \cite{K-talk}
by one of the authors of \cite{KrZu10}.
On the negative side, contrary to \cite{KrZu10},
we need to impose the \textit{additional} hypothesis
that the second derivative $\ddot\theta$ exists and is bounded.
However, it seems that one could get rid of this technical
assumption by adapting an approximation technique
recently proposed in \cite{KrSe}, \cite{KS}. While preparing this paper we learned about a recent result \cite{KrSe}, 
where the norm-resolvent convergence in the limit of thin quantum wave-guides is proved under certain "mild" regularity conditions. 
The key step is different to our method and is based on the Steklov approximation.

The paper is organized as follows. In the next Section 2
we recall the origin of the operator \eqref{self-Hamiltonian}
in more details and formulate our main Theorem.
The proof essentially consists of three steps
and is correspondingly presented in Section 3.
The paper is concluded in Section 4
by mentioning a more general model.

\section{Set up and the main Theorem}
Let $\Omega_0:=\mathbb{R}\times \omega$ be a straight tube
with the main axis $\mathbb{R}$ and a non-circular
cross section, which is a bounded connected open set $\omega\subset \mathbb{R}^2$. Let $\Omega_\theta$ denote the
corresponding \textit{locally twisted} tube with the same main axis. This means that we allow $\omega$ to rotate with
variation of the coordinate $x_1$ along the main axis $\mathbb{R}$ on (non-constant) angle $\theta: x_1\mapsto \theta(x_1)$,
and we assume that this twisting is \textit{smooth} and local, i.e. the derivative $\dot{\theta}(x_1)$ is a $C^{1}$-smooth
function with \textit{compact support} in $\mathbb{R}$, see Figure \ref{Fig1}.
%
%
With our choice of the main axis, for $x:=(x_1,x_2,x_3) \in \mathbb{R}^3$
we refer to $x_1$ as the `` longitudinal" and to $x'=(x_2,x_3)$ as the `` transverse" coordinates
in the tubes $\Omega_0$ and $\Omega_\theta$. Then
transition from the straight to the twisted tube
is the mapping $\mathcal{L}_{\theta}: \Omega_0 \rightarrow \Omega_\theta$
defined explicitly by the function
\begin{equation*}
\mathcal{L}_{\theta}(x):= (x_1, x_2 \cos\theta(x_1) + x_3 \sin\theta(x_1), - x_2 \sin\theta(x_1) + x_3 \cos\theta(x_1)) \ .
\end{equation*}
We consider in $\Omega_0$ and in $\Omega_\theta$, i.e. in the spaces $L^2(\Omega_0)$ and $L^2(\Omega_\theta)$, the
(\textit{minus}) Dirichlet Laplacians. We denote them respectively by $-\Delta^{\Omega_0}_D$  and $-\Delta^{\Omega_\theta}_D$.

For the case of the straight tube twisting $\mathcal{L}_{\theta}$ there is an $x_1$-dependent local rotation of
coordinates that maps the twisted tube $\Omega_\theta$ into the straight one $\Omega_0$. Let
$V_\theta: L^2(\Omega_\theta)\rightarrow L^2(\Omega_0)$ denote the unitary representation of this mapping:
$V_\theta\psi := \psi \circ \mathcal{L}_\theta$. Then
the corresponding unitary transformation of the twisted Dirichlet Laplacians $-\Delta^{\Omega_\theta}_D$
takes the form
%
%
\cite{Kr08}, \cite{KrZu10}:
\begin{equation}\label{straight}
H_{\theta}:=V_\theta(-\Delta^{\Omega_\theta}_D)V_\theta^{-1}=
-(\partial_1-\dot{\theta}\partial_\tau)^2-\Delta_D^\omega \ ,\quad \mbox{with}\quad \mathrm{dom}(H_{\theta}) \subset L^2{(\Omega_0)} \ .
\end{equation}
The quadratic form associated to self-adjoint operator $H_\theta$ is
\begin{equation}\label{straight-QForm}
\mathcal{Q}_\theta[\psi]:=
||\partial_1\psi-\dot{\theta}\partial_\tau \psi||^2_{L^2(\Omega_0)}+
||\nabla'\psi||^2_{L^2(\Omega_0)} \ ,
\end{equation}
with domain $\mathrm{dom}(\mathcal{Q}_\theta)= W^{1,2}_{0}{(\Omega_0)}$, which is Sobolev space in $L^2{(\Omega_0)}$.
Here we denote by $\nabla':=(\partial_2,\partial_3)$ the transverse gradient in $\omega$, i.e.
$\Delta_D^\omega:=(\partial_2^2 + \partial_3^2)_D$
stays for {Dirichlet Laplacian} operator in the space $L^2{(\omega)}$,
%
%
corresponding to \textit{cross-section} $\omega$, and the operator
\begin{equation*}
\partial_\tau:=\tau \cdot \nabla'=x_3\partial_2-x_2\partial_3 \ , \quad
 \mbox{for vector}\quad \tau=(x_3,-x_2) \ ,
\end{equation*}
is the angular-derivative in $\mathbb{R}^2 \supset \omega$.

To describe the limit of (\ref{straight}) for \textit{simultaneous} wave-guide diameter and twisting supports
shrinking, we use instead of the self-similar parametrization
(see \cite{KrZu10}, Ch.1.2, IV) the
following family of \textit{scaled} operators.

We denote by $U_\varepsilon$ the unitary transformation acting as $(U_\varepsilon \psi)(x):=
\sqrt{\varepsilon}\ \psi(\varepsilon \ x_1,x_2,x_3)$, for $\varepsilon >0$, and we introduce the family of
\textit{scaled} operators $\hat{H}_{\varepsilon, \theta}$:
\begin{equation}\label{scaled-eps}
\hat{H}_{\varepsilon, \theta} =
\varepsilon^2 U_\varepsilon^\ast {H_\theta} U_ \varepsilon=
-(\partial_1-\sigma_\varepsilon\partial_\tau)^2-
\frac{1}{\varepsilon^2} \ \Delta_D^\omega \ , \quad \mbox{in}\quad L^2{(\Omega_0)}.
\end{equation}
Here $\hat{H}_{\varepsilon, \theta}$ is associated with the quadratic form
\begin{equation}\label{scaled-eps-qform}
\mathcal{\widehat{Q}}_{\varepsilon, \theta}[\psi]:=||\partial_1\psi-
\sigma_\varepsilon\partial_\tau \psi||^2_{L^2(\Omega_0)}+
\frac{1}{\varepsilon^2}||\nabla'\psi||^2_{L^2(\Omega_0)} \ ,
\end{equation}
with domain $\mathrm{dom}(\mathcal{\widehat{Q}}_{\varepsilon, \theta})= W^{1,2}_{0}{(\Omega_0)}$.
Here $\sigma_\varepsilon(\cdot): = {\varepsilon}^{-1} \ \dot{\theta}({\cdot}/{\varepsilon})$, i.e. support
of twisting decreases, when $\varepsilon \rightarrow 0$, and $\sigma_\varepsilon(\cdot)$ becomes \textit{singular} in
cross-section $\{x_1 =0\}\times \omega$. To appreciate this singularity notice that in distributional sense
$\lim_{\varepsilon \rightarrow 0}\sigma_\varepsilon(\cdot)$ exists and coincides with
$(\theta(+\infty) - \theta(-\infty)) \ \delta_{0}(\cdot)$, where $\delta_{0}(\cdot)$ is the Dirac symbol with support at
$x_1 = 0$. Below we are dealing with even stronger singularity due to $\sigma^{2}_\varepsilon(\cdot)$.

Let $E_1 >0$ denote the first eigenvalue of the operator $(- \Delta_D^\omega)$ in the cross-section $\omega$. Then by
virtue of (\ref{scaled-eps-qform}) the value $E_1/\varepsilon^2$ is the lower bound (and the spectral \textit{infimum})
of the operator (\ref{scaled-eps}). This bound increases for $\varepsilon \rightarrow 0$ with the rate corresponding to
geometrical shrinking of the cross-section.

Following the strategy of \cite{K-talk}-\cite{KrZu11} the next step is to investigate the operator (\ref{straight})
in a "natural" weighted Sobolev space $W^{1,2}_{0}(\Omega_0, K(x) dx)$ corresponding to
$\mathcal{H}_{k}:=L^2(\Omega_0, K^{k}(x) dx)$ for $k=1$, where $K(x) = \exp (x_{1}^2/4)$, see \cite{KrZu10} Ch.5.3.
The advantage of this approach is that in the space $\mathcal{H}_{1}$ (instead of $\mathcal{H}_{0}$) the corresponding
operator (\ref{scaled-eps}) has a compact resolvent. Indeed, let the transformation
$\mathcal{U}_{K}: \mathcal{H}_{1} \rightarrow \mathcal{H}_{0}$ is defined by
\begin{equation*}
(\mathcal{U}_{K} \phi)(x):= (K^{1/2} \phi)(x)= e^{x_{1}^2/8} \ \phi(x_{1},x_{2},x_{3}) \ ,
\end{equation*}
and let ${H}_{\varepsilon, \theta}: = \mathcal{U}_{K} \ \hat{H}_{\varepsilon, \theta} \ \mathcal{U}_{K}^{-1}$.
Then operator (\ref{scaled-eps}) is unitary equivalent to
\begin{equation}\label{Oper H_0-H_1}
H_{\varepsilon, \theta}=
-(\partial_1-\sigma_\varepsilon\partial_\tau)^2-
\frac{1}{\varepsilon^2} \ \Delta_D^\omega+\frac{x^2_1}{16},\quad \mbox{in}\quad \mathcal{H}_{0}= L^2{(\Omega_0)} \ ,
\end{equation}
which is self-adjoint operator associated to the quadratic form
\begin{equation}\label{Q-form H_0-H_1}
\mathcal{Q}_{\varepsilon, \theta}[\psi]:=||\partial_1\psi-
\sigma_\varepsilon\partial_\tau \psi||^2_{L^2(\Omega_0)}+
\frac{1}{\varepsilon^2}||\nabla'\psi||^2_{L^2(\Omega_0)}
+\frac{1}{16}||x_1\psi||^2_{L^2(\Omega_0)} \
\end{equation}
with \textit{domain} in the \textit{weighted space} $W^{1,2}_{0}(\Omega_0, K(x) dx)$.
Therefore, the harmonic potential in direction $x_1$, together with Dirichlet Laplacian $\Delta_D^\omega$ in cross-section
$\omega$ with the discrete spectrum
$\mathrm{Sp}(-\Delta_D^\omega)= \{E_1 < E_2 \leq E_3 \leq \ldots \}$,
make the total spectrum $\mathrm{Sp}(H_{\varepsilon, \theta})$ of the operator (\ref{Oper H_0-H_1}) \textit{pure point} and increasing to
infinity. This bolsters the claim that the resolvent of (\ref{Oper H_0-H_1}) is compact.

Notice that shrinking ($\varepsilon \rightarrow 0$) of the cross-section implies via transversal operator $(- \Delta_D^\omega/{\varepsilon^2})$
the shift of $E_n/\varepsilon^2 \rightarrow\infty $ and of the whole spectrum $\mathrm{Sp}(H_{\varepsilon, \theta})$ to \textit{infinity}.
Hence, to make a sense of a resolvent limit for (\ref{Oper H_0-H_1}) one has to study the shifted resolvent
\begin{equation}\label{shift-res}
R_{(E_1/\varepsilon^2-1)}(H_{\varepsilon, \theta}):= (H_{\varepsilon, \theta} - E_1/\varepsilon^2  + 1)^{-1} \ ,
\end{equation}
which is well-defined since by (\ref{Q-form H_0-H_1}) one has $H_{\varepsilon, \theta}- E_1/\varepsilon^2 + 1 > 1$
uniformly in $\varepsilon > 0$.

To proceed to formulation of our main result we single out from (\ref{Oper H_0-H_1}) the \textit{one-dimensional}
harmonic oscillator operator $h_0 > 0$:
\begin{equation}\label{h0}
h_0: =-\frac{d^2}{dx_1^2}+\frac{x_1^2}{16} \ \ , \quad \mbox{in}\quad L^2(\mathbb{R}) \ ,
\end{equation}
and introduce the operator $h_0^D \geq h_0$ defined as $h_0$, but with \textit{Dirichlet boundary condition} at $x_1=0$:
\begin{equation}\label{dom-h-D}
\mathrm{dom}((h_0^D)^{1/2}):= \{u\in \mathrm{dom}((h_0)^{1/2}): u(x_1 =0)=0 \} \ .
\end{equation}

The aim of the present paper is to compare the \textit{shifted} operator
$H_{\varepsilon, \theta} -  E_1/\varepsilon^2$ and $h_0^D$ in the \textit{norm-resolvent} sense. This makes a
difference between our result and  \cite{K-talk}-\cite{KrZu11}, where the convergence of these operators for
$\varepsilon \rightarrow 0$ was established in the \textit{strong-resolvent} sense.

Since operators $H_{\varepsilon, \theta} -  E_1/\varepsilon^2$ and $h_0^D $ act in \textit{different} spaces we have
to elucidate the above statement decomposing $\mathcal{H}_{0} = L^2(\Omega_0)$ into orthogonal sum:
\begin{equation}\label{decomp}
\mathcal{H}_{0} = \mathfrak{H}_{1} \oplus \mathfrak{H}_{1}^{\bot} \ .
\end{equation}
Here
$\mathfrak{H}_{1}:= \{u \otimes \mathcal{J}_1 : u(x_1) \in L^2(\mathbb{R}), \ \mathcal{J}_1 (x'): (- \Delta_D^\omega) \mathcal{J}_1
= E_1 \mathcal{J}_1 \ , x' =(x_2,x_3) \}$. With this decomposition we obtain
\begin{eqnarray}\label{decomp-ident-1}
&&(H_{\varepsilon, \theta} -  E_1/\varepsilon^2 +1) \ u \otimes \mathcal{J}_1 =\\
&& [-(\partial_1 \otimes I -  \sigma_\varepsilon \otimes\partial_\tau)^2- I \otimes \frac{1}{\varepsilon^2} \
(\Delta_D^\omega + E_1 -1) + \frac{x^2_1}{16}\otimes I ]\ u \otimes \mathcal{J}_1 = \nonumber \\
&&[-(\partial_1 \otimes I -  \sigma_\varepsilon \otimes\partial_\tau)^2 +
({x^2_1}/{16}+1)\otimes I] \ u \otimes \mathcal{J}_1  \ , \nonumber
\end{eqnarray}
and the estimate on $\mathfrak{H}_{1}^{\bot}$ from below:
\begin{eqnarray}\label{decomp-ident-2}
&&(v \otimes \mathcal{J}_{n > 1}, ((E_n - E_1)/{\varepsilon^2} + 1) \
v \otimes \mathcal{J}_{n > 1})_{\mathfrak{H}_{1}^{\bot}} \leq \\
&&(v \otimes \mathcal{J}_{n > 1},(H_{\varepsilon, \theta}  -  E_1/\varepsilon^2 + 1) \
v \otimes \mathcal{J}_{n > 1})_{\mathfrak{H}_{1}^{\bot}}  \  \ . \nonumber
\end{eqnarray}
This decomposition allows us also to construct a suitable extension of the resolvent
$\widehat{R}_{(z=-1)}(h_0^D):=(h_0^D + 1)^{-1}$ (originally defined on $L^2(\mathbb{R})$) to the whole space $\mathcal{H}_{0}$.
Below we denote this \textit{extension} by $R_{z}(h_0^D)$.

To this end notice that operator $(h_0^D + 1)\otimes I$ is invertible in $\mathfrak{H}_{1}$. Hence, we can extend
this inversion by \textit{zero} operator $0^{\bot}$ on $\mathfrak{H}_{1}^{\bot}$  and define :
\begin{equation}\label{resolv-h-D}
R_{(z=-1)}(h_0^D): = (h_0^D + 1)^{-1}\otimes I \oplus 0^\perp  \ .
\end{equation}
This extension is evidently motivated by (\ref{decomp}) and (\ref{decomp-ident-2}) saying that for
$\varepsilon \rightarrow 0$ the resolvent (\ref{shift-res}) converges to the \textit{zero} operator $0^{\bot}$ on
$\mathfrak{H}_{1}^{\bot}$.

Now we are in position to formulate our main result.
\begin{theorem}\label{main-theorem}
Let $\Omega_\theta$ be a twisted tube with $\dot{\theta}\in C^1_0(\mathbb{R})$ and with a bounded
$\ddot{\theta}$. Then,
\begin{equation}\label{main-theorem eq}
\lim_{\varepsilon \rightarrow 0} ||(H_{\varepsilon, \theta} - E_1/\varepsilon^2 +1)^{-1}-
\left[(h_0^D+1)^{-1}\otimes I \oplus 0^\perp\right]|| = 0 \ ,
\end{equation}
in the operator norm of the space $\mathcal{H}_{0}= L^2(\Omega_{0})$.
\end{theorem}
\begin{remark}\label{Rem1}
Using decomposition (\ref{decomp}) we split the proof of the Theorem into several steps.
To this end we introduce in $\mathcal{H}_{0}= \mathfrak{H}_{1} \oplus \mathfrak{H}_{1}^{\bot}$ the \textit{intermediate}
operator:
\begin{eqnarray}\nonumber
H_0^\varepsilon&:=& [-(\partial_1^2 + \frac{x^2_1}{16} + C_\omega \sigma^{2}_{\varepsilon} )\otimes I
+ I \otimes (-\Delta_D^\omega)/{\varepsilon^2} ] \\
&=:&h_{\varepsilon}\otimes I + I \otimes \frac{1}{\varepsilon^2}\bigoplus_{n=1}^{\infty}{E_n} \, P_{n} \ ,
\label{interm-oper}
\end{eqnarray}
where $C_\omega:=||\partial_\tau \mathcal{J}_1||^2_{L^2(\omega)}$
and $P_n:=(\mathcal{J}_n, \cdot )_{L^2(\omega)} \ \mathcal{J}_n$
are orthogonal projectors on the transversal modes $\mathcal{J}_n,$ $n=1,2,3,\ldots$ . We denote by
$R_{z}(H_0^\varepsilon):= (H_0^\varepsilon - z \ I \otimes I)^{-1}$ its resolvent at the point $z$ in the
resolvent set and we denote by $\mathcal{Q}_0^\varepsilon$ the sesquilinear form associated with $H_0^\varepsilon.$
\end{remark}
\begin{remark}\label{Rem2}
Since operator $(h_{\varepsilon} +1)\otimes I$ is invertible
in $\mathfrak{H}_{1}$, then similarly to (\ref{resolv-h-D})
we define the resolvent:
%
%
\begin{equation}\label{resolv-h-varepsilon}
R_{(z=-1)}(h_{\varepsilon}): = (h_{\varepsilon} + 1)^{-1}\otimes I \oplus 0^\perp  \ .
\end{equation}
\end{remark}

Notice that by (\ref{decomp-ident-1}) and (\ref{interm-oper}) the difference of resolvents:
\begin{equation}\label{diff-res}
R_{(E_1/\varepsilon^2-1)}(H_{\varepsilon, \theta}) - R_{(E_1/\varepsilon^2-1)}(H_0^\varepsilon)=
R_{(E_1/\varepsilon^2-1)}(H_{\varepsilon, \theta}) \ (H_0^\varepsilon - H_{\varepsilon, \theta})\
R_{(E_1/\varepsilon^2-1)}(H_0^\varepsilon) \ ,
\end{equation}
is finite on $\mathfrak{H}_{1}$ and tends to zero (for $\varepsilon \rightarrow 0$) on $\mathfrak{H}_{1}^{\bot}$ ,
cf (\ref{decomp-ident-2}). Hence, the \textit{first step} is to compare the operators (\ref{Oper H_0-H_1}) and
(\ref{interm-oper}).

Since similar to (\ref{decomp-ident-2}) the resolvent $R_{(E_1/\varepsilon^2-1)}(H_0^\varepsilon)$
converges for $\varepsilon \rightarrow 0$ to the \textit{zero} operator $0^{\bot}$ on
$\mathfrak{H}_{1}^{\bot}$, our \textit{second step} is to compare (in the proper sense) the total operator (\ref{interm-oper})
with operator $h_{\varepsilon}\otimes I$ acting in $\mathfrak{H}_{1}$ and defined by the resolvent (\ref{resolv-h-varepsilon})
as "infinity" in the complement subspace $\mathfrak{H}_{1}^{\bot}$.

The \textit{third step} is to prove the norm-resolvent convergence of operators $h_{\varepsilon}$ and
$h_0^D$, which is reduced to analysis in $L^2(\mathbb{R})$ and to technique due to \cite{BGW}.
\noindent
\section{{Proofs}}

As it is mentioned at the end of Section 2, the proof of Theorem \ref{main-theorem} is divided into
three steps and to prove this theorem, we use the intermediate operator \eqref{interm-oper} and the operator
$h_{\varepsilon}\otimes I$ via definition (\ref{resolv-h-varepsilon}).
We insert the corresponding resolvents $R_{(E_1/\varepsilon^2-1)}(H_0^\varepsilon)$ and $R_{(z=-1)}(h_{\varepsilon})$ into the
limit \eqref{main-theorem eq} in the following way:
\begin{eqnarray*}
&&||R_{(E_1/\varepsilon^2-1)}(H_{\varepsilon, \theta})-R_{(E_1/\varepsilon^2-1)}(H_0^{\varepsilon})
+ R_{(E_1/\varepsilon^2-1)}(H_0^{\varepsilon})- R_{(z=-1)}(h_{\varepsilon})+\\
&&R_{(z=-1)}(h_{\varepsilon}) - R_{(z=-1)}(h_0^D)||
\end{eqnarray*}
{Hence the operator norm of the resolvent difference in \eqref{main-theorem eq} is bounded by the three terms:}
\begin{eqnarray}\nonumber
&&||R_{(E_1/\varepsilon^2-1)}(H_{\varepsilon, \theta})-R_{(E_1/\varepsilon^2-1)}(H_0^{\varepsilon})||
+||R_{(E_1/\varepsilon^2-1)}(H_0^{\varepsilon})- R_{(z=-1)}(h_{\varepsilon})|| + \\
&&\|R_{(z=-1)(h_{\varepsilon})} - R_{(z=-1)}(h_0^D)|| . \label{3-terms}
\end{eqnarray}
We estimate them separately in the following three steps below.

\subsection{Step one}
{First we estimate the operator norm of the difference (\ref{diff-res}). To this end we compare
the quadratic forms $\mathcal{Q}_{\varepsilon, \theta}$ (see \eqref{Q-form H_0-H_1}) and $\mathcal{Q}_0^\varepsilon$
and to show that the difference $m_\varepsilon:=\mathcal{Q}_0^\varepsilon-\mathcal{Q}_{\varepsilon, \theta}$ goes to
zero as $\varepsilon$ goes to zero. This would mean that the problem of approximation is reduced now to analysis of
the intermediate operator (\ref{interm-oper}) or the form $\mathcal{Q}_0^\varepsilon$}.

{For this purpose we denote by $\phi, \psi \in \mathcal{H}_0 = L^2(\Omega_0)$ the solutions of equations:}
\begin{equation}\label{Eqs}
F=(H_{\varepsilon, \theta}- E_1/\varepsilon^2 + 1)\phi,\quad G=(H_0^{\varepsilon}- E_1/\varepsilon^2 + 1)\psi,
\quad F,G \in \mathcal{H}_0 \ .
\end{equation}
\noindent
Then we obtain for the difference (\ref{diff-res}) the representation:
\begin{equation}\label{repr-m}
(F,\left(R_{(E_1/\varepsilon^2-1)}(H_{\varepsilon, \theta})-R_{(E_1/\varepsilon^2-1)}(H_0^{\varepsilon})\right)G)=
\mathcal{Q}_0^\varepsilon-\mathcal{Q}_{\varepsilon, \theta}= m_\varepsilon(\phi, \psi) \ ,
\end{equation}
where, the sesquilinear form $m_\varepsilon(\phi, \psi)$ is explicitly given by
\begin{equation}\label{sesquilinear form}
m_\varepsilon(\phi,\psi)=(\phi,C_\omega\sigma_\varepsilon^2\psi)+
(\partial_1\phi,\sigma_\varepsilon\partial_\tau \psi)+
(\sigma_\varepsilon\partial_\tau \phi,\partial_1 \psi)-
(\partial_\tau \phi, \sigma^2_\varepsilon \partial_\tau \psi).
\end{equation}
\begin{lemma}\label{lemma01}
For $\varepsilon \rightarrow 0$ the sesquilinear form \eqref{sesquilinear form} can be estimated as:
\begin{equation}\label{m<}
|m_\varepsilon(\phi,\psi)|\leq \ \varepsilon \ C_m  \, ||F||_{\mathcal{H}_0}||G||_{\mathcal{H}_0}, \quad F,G \in \mathcal{H}_0 \ ,
\end{equation}
for a certain constant $C_m > 0$ and for solutions $\phi,\psi$ of (\ref{Eqs}).
\end{lemma}
\noindent
\textbf{Proof}. Following decomposition (\ref{decomp}) we represent the functions
$\psi, \phi \in \mathcal{H}_0$ as $\psi=\psi_1 \oplus \psi_1^\perp$ and
$\phi=\phi_1 \oplus \phi_1^\perp$, where $\psi_1, \phi_1 \in {\mathfrak{H}}_{1}$  and
$\psi_1^\perp, \phi_1^\perp \in {\mathfrak{H}}_{1}^{\bot}$.  Then we obtain
\begin{equation}\label{m0}
m_\varepsilon(\phi, \psi)=m_\varepsilon(\phi_1,\psi_1)+m_\varepsilon(\phi_1^\perp,\psi_1^\perp)
+m_\varepsilon(\phi_1, \psi_1^\perp)+m_\varepsilon(\phi_1^\perp,\psi_1).
\end{equation}
\noindent

First, we show that $m_\varepsilon(\phi_1, \psi_1)=O(\varepsilon)$ and
$m_\varepsilon(\phi_1^\perp,\psi_1^\perp)=O(\varepsilon)$. To this end, we use (\ref{sesquilinear form}) to write
explicitly
\begin{equation}\label{m1}
m_\varepsilon(\phi_1, \psi_1)=(\phi_1,C_\omega\sigma_\varepsilon^2\psi_1)-
(\partial_\tau \phi_1, \sigma^2_\varepsilon \partial_\tau \psi_1)+
(\partial_1\phi_1,\sigma_\varepsilon\partial_\tau \psi_1)+
(\sigma_\varepsilon\partial_\tau \phi_1,\partial_1 \psi_1) \ .
\end{equation}
\noindent
To compute the first two terms in the right-hand side of (\ref{m1}) we use definition of $C_\omega$ and the fact that
$\phi_1= u(x_1)\otimes\mathcal{J}_1(x')$ and $\psi_1= v(x_1)\otimes\mathcal{J}_1(x')$, where $\mathcal{J}_1$ is
normalized to one. Then one gets that these terms vanish:
\begin{eqnarray}
&& (\phi_1,C_\omega\sigma_\varepsilon^2\psi_1) -
(\partial_\tau \phi_1, \sigma^2_\varepsilon \partial_\tau \psi_1) = \nonumber\\
&&{C_\omega \int_{\mathbb{R}}\sigma^2_\varepsilon(x_1)\overline{u}(x_1) v(x_1)dx_1
\int_\omega |\mathcal{J}_1(x')|^2 dx' -
\int_{\mathbb{R}}\sigma^2_\varepsilon(x_1)\overline{u}(x_1) v(x_1)dx_1 \
||\partial_\tau \mathcal{J}_1||^2_{L^2(\omega)} \ . }\nonumber\\ \label{proc-1}
\end{eqnarray}
\noindent
{To estimate the last two terms in the right-hand side of (\ref{m1}) we use equations (\ref{Eqs}).
In particular they imply that $\sigma^2_\varepsilon(x_1) {u}(x_1) \in L^2(\mathbb{R})$, or :}
\begin{equation}\label{Est1}
\int_{\mathbb{R}} \frac{1}{\varepsilon^4}(\dot{\theta}(x_1/\varepsilon))^{4} |u(x_1)|^2 dx_1 =
\frac{1}{\varepsilon^3}\int_{\mathbb{R}} (\dot{\theta}(y))^{4} |u(\varepsilon y)|^2 dy < C_{u} \ .
\end{equation}
{By conditions on $\dot{\theta}$ this means that solutions of equations (\ref{Eqs})
have asymptotic}
\begin{equation}\label{asympt}
u(\varepsilon y) = O(\varepsilon^{3/2}) \ \ {\rm{for}} \ \ \varepsilon \rightarrow 0  \ \
{\rm{and}} \ \ y\in K \ ,
\end{equation}
for any compact $K \subset \mathbb{R}$. Then to estimate the third term in the right-hand side of (\ref{m1}) we use (\ref{asympt}).
This gives:
\begin{eqnarray}\label{Est2}
&&|(\partial_1\phi_1,\sigma_\varepsilon\partial_\tau \psi_1)| =
\left|\int_{\mathbb{R}} \partial_1\overline{u}(x_1) \frac{1}{\varepsilon}\ \dot{\theta}(x_1/\varepsilon)\ v(x_1) dx_1
\int_\omega \mathcal{J}_1(x')\partial_\tau \mathcal{J}_1(x') dx'\right| \leq  \\
&& C_\omega  \|\partial_1{u}\|_{L^2(\mathbb{R})}
\left\{\int_{\mathbb{R}} \frac{1}{\varepsilon}(\dot{\theta}(y))^2|v(\varepsilon y)|^2 dy\right\}^{1/2} \leq
O(\varepsilon) \ C_\omega  \|\partial_1{u}\|_{L^2(\mathbb{R})}
\left\{\int_{\mathbb{R}} (\dot{\theta}(y))^2 dy\right\}^{1/2}. \nonumber
\end{eqnarray}
{Since by (\ref{Eqs}) $\partial_1{u}\in L^2(\mathbb{R})$, the inequality (\ref{Est2}) implies the estimate
$|(\partial_1\phi_1,\sigma_\varepsilon\partial_\tau \psi_1)| = O(\varepsilon)$. Similarly one obtain the
estimate $(\sigma_\varepsilon\partial_\tau \phi_1,\partial_1 \psi_1)= O(\varepsilon)$, that yields
$m_\varepsilon(\phi_1, \psi_1)= O(\varepsilon)$.}

We can show that $m_\varepsilon(\phi_1^\perp, \psi_1^\perp)=O(\varepsilon)$ by similar calculations.
Indeed, we have representation:
$$
m_\varepsilon(\phi_1^\perp, \psi_1^\perp)=(\phi_1^\perp,C_\omega\sigma_\varepsilon^2\psi_1^\perp)
-(\partial_\tau \phi_1^\perp, \sigma^2_\varepsilon \partial_\tau \psi_1^\perp)+
(\partial_1\phi_1^\perp,\sigma_\varepsilon\partial_\tau \psi_1^\perp)+
(\sigma_\varepsilon\partial_\tau \phi_1^\perp,\partial_1 \psi_1^\perp).
$$
\noindent
Then in a complete similarity with (\ref{m1}) one obtains that the terms $|(\partial_1\phi_1^\perp,\sigma_\varepsilon\partial_\tau
\psi_1^\perp)|$ and $|(\sigma_\varepsilon\partial_\tau \phi_1^\perp,\partial_1 \psi_1^\perp)|$ are of order $\varepsilon$ and that
$$
(\phi_1^\perp,C_\omega\sigma_\varepsilon^2\psi_1^\perp)-
(\partial_\tau \phi_1^\perp, \sigma^2_\varepsilon \partial_\tau \psi_1^\perp)
=0 \ .
$$
\noindent

Now let us estimate the term
\begin{equation}\label{m3}
m_\varepsilon(\phi_1,\psi_1^\perp)=
(\sigma_\varepsilon \partial_\tau \phi_1,\partial_1 \psi^\perp_1)
-(\partial_\tau\phi_1,\sigma_\varepsilon^2 \partial_\tau \psi_1^\perp)+
(\partial_1\phi_1,\sigma_\varepsilon \partial_\tau \psi_1^\perp)+
(C_\omega\sigma^2_\varepsilon \phi_1,\psi^\perp_1).
\end{equation}
\noindent
Since $\phi_1= u\otimes\mathcal{J}_1$ and $\psi_1^\perp$ belongs
to the linear envelope of $\{v \otimes\mathcal{J}_{n}\}_{n=2}^\infty$,
%
%
to estimate the first term in \eqref{m3} we consider:
\begin{equation}\label{estimate-2}
(\sigma_\varepsilon \partial_\tau \phi_1,\partial_1 \psi_1^\perp)=
\int_\mathbb{R}\frac{1}{\varepsilon}\ \dot{\theta}(x_1/\varepsilon)\ u(x_1) \partial_1 v(x_1) dx_1 \int_\omega \partial_\tau
\mathcal{J}_1(x')\{\mathcal{J}_{n}\}_{n=2}^\infty(x')dx' \ .
\end{equation}
Notice that integral (\ref{estimate-2}) coincides (up to simple modifications) with the integral in (\ref{Est2}).
Therefore, it has the same estimate $O(\varepsilon)$.
Similarly we obtain for the third term in \eqref{m3} the representation:
\begin{equation}\label{estimate-3}
(\partial_1\phi_1,\sigma_\varepsilon \partial_\tau \psi_1^\perp)=
\int_\mathbb{R}\partial_1 u(x_1) \frac{1}{\varepsilon}\ \dot{\theta}(x_1/\varepsilon) v(x_1) dx_1 \int_\omega
\mathcal{J}_1(x')\partial_\tau\{\mathcal{J}_{n}\}_{n=2}^\infty(x')dx' \ ,
\end{equation}
\noindent
which implies that this term is also of the order $O(\varepsilon)$.
\noindent
To estimate the term $(\partial_\tau \phi_1,\sigma_\varepsilon^2 \partial_\tau \psi^\perp_1)$,
we use the following inequalities:
\begin{eqnarray}
|(\partial_\tau \phi_1,\sigma_\varepsilon^2 \partial_\tau \psi^\perp_1)|&=&
\left|\int_\mathbb{R} u(x_1)\, \frac{1}{\varepsilon^2}(\dot{\theta}({x_1}/{\varepsilon}))^2 \, v_1(x_1)dx_1 \,
\int_{\omega} \partial_\tau\mathcal{J}_1(x') \partial_\tau \mathcal{J}_{s>1}(x')\,dx'\right|  \nonumber\\
&\leq &C_\omega\left\{\int_{\mathbb{R}}\, \frac{1}{\varepsilon^2}(\dot{\theta}({x_1}/{\varepsilon}))^2 \,|u(x_1)|^{2} dx_1\right\}^{1/2}
\left\{\int_{\mathbb{R}}\, \frac{1}{\varepsilon^2}(\dot{\theta}({x_1}/{\varepsilon}))^2 \,|v(x_1)|^{2} dx_1\right\}^{1/2}  \nonumber\\
&= & C_\omega\left\{\int_{\mathbb{R}}\, \frac{1}{\varepsilon}(\dot{\theta}(y))^2 \,|u(\varepsilon y)|^{2} dy \right\}^{1/2}
\left\{\int_{\mathbb{R}}\, \frac{1}{\varepsilon}(\dot{\theta}(y))^2 \,|v(\varepsilon y)|^{2} dy\right\}^{1/2}  \nonumber\\
&\leq& O(\varepsilon^2) \ C_\omega  \ , \label{proceed1}
\end{eqnarray}
where the last asymptotic follows from (\ref{Est1}) and (\ref{asympt}). Finally, since $\phi_1$ and $\psi^\perp_1$ belong to
orthogonal subspaces we obtain for the last term $(C_\omega\sigma_\varepsilon^2\phi_1,\psi_1^\perp)=0.$

Note that the estimate of the term $m(\phi_1^\perp,\psi_1)$ is identical to $m(\phi_1,\psi_1^\perp)$.
Therefore, summarizing  \eqref{Est2}, \eqref{estimate-2}, \eqref{estimate-3}, and \eqref{proceed1}, we obtain
the estimate of the form (\ref{m0}) for solutions of (\ref{Eqs}) by $O(\varepsilon)$. Since
equations (\ref{Eqs}) yield the estimate of $\phi,\psi$ by norms $||F||_{\mathcal{H}_0}, ||G||_{\mathcal{H}_0}$, one gets
(\ref{m<}). So, the proof of Lemma \ref{lemma01} is completed.\quad \QED
\noindent
\begin{remark}\label{rem-proceed}
By (\ref{repr-m}) and (\ref{m<}) we obtain the rate of the operator-norm convergence for the difference of resolvents
(\ref{diff-res}):
\begin{equation}\label{norm-res}
\|R_{(E_1/\varepsilon^2-1)}(H_{\varepsilon, \theta})-R_{(E_1/\varepsilon^2-1)}(H_0^{\varepsilon})\| \leq
\varepsilon \ C_m  \ .
\end{equation}
\end{remark}
\noindent

\subsection{Step two}
By virtue of definitions (\ref{interm-oper}) and (\ref{resolv-h-varepsilon}) we obtain
\begin{eqnarray}\label{step2-repr1}
&&\Lambda_{\varepsilon}:= R_{(E_1/\varepsilon^2-1)}(H_0^\varepsilon) - R_{(z=-1)}(h_{\varepsilon}) = \\
&&[(h_{\varepsilon}+1)\otimes I + I \otimes \frac{1}{\varepsilon^2}\bigoplus_{n=2}^{\infty}(E_n - E_1)  \,
P_{n}]^{-1}-[(h_{\varepsilon} + 1)^{-1}\otimes I \oplus 0^\perp] \ .\nonumber
\end{eqnarray}
Since $P_{n>1}: \mathfrak{H}_{1} \rightarrow 0$, one gets $\Lambda_{\varepsilon}\phi = 0$ for $\phi\in\mathfrak{H}_{1}$. On the hand
for $\phi^{\bot}\in\mathfrak{H}_{1}^{\bot}$ we have:
\begin{equation}\label{step2-repr2}
\Lambda_{\varepsilon} \, \phi^{\bot} = [I \otimes \frac{1}{\varepsilon^2}\bigoplus_{n=2}^{\infty}(E_n - E_1)\, P_{n}]^{-1} \, \phi^{\bot} \ .
\end{equation}
Therefore, for the second term in (\ref{3-terms}) we obtain the estimate
\begin{equation}\label{est-step2}
||R_{(E_1/\varepsilon^2-1)}(H_0^{\varepsilon})-
R_{(z=-1)}(h_{\varepsilon})||\leq \varepsilon^2/(E_2 - E_1) \ .
\end{equation}
\subsection{Step three}
Recall the definition \eqref{interm-oper} of the intermediate operator
$$
h_\varepsilon=-\partial_1^2+\frac{x_1^2}{16}+C_\omega \sigma^2_\varepsilon
$$
\noindent
and recall that the operator $h_0$ is the operator $-\partial_1^2+\frac{x_1^2}{16}$ define on $L^2(\mathbb{R})$
while $h_0^D$ is the analoguous operator plus a Dirichlet boundary condition at the origin.
Let us denote
$$
R_{k^2}(h_\varepsilon):=(h_\varepsilon-k^2)^{-1},\quad r_{k^2}(h_0^D):=
(h_0^D-k^2)^{-1}, \quad k^2 \notin \sigma (h_\varepsilon)
$$
\noindent
The third step consists in showing the following lemma:
\begin{lemma}\label{lemmeprincipal}
Let $h_\varepsilon$ $h_0$ and $h_0^D$ being the operators on $L^2(\mathbb{R})$
described as above (see \eqref{h0}). Let us denote
$ R(h_0):=(h_\varepsilon-k^2)^{-1},\quad R(h_0^D):=(h_0^D-k^2)^{-1}.$ Then we get
$$
\lim_{\varepsilon \rightarrow 0} ||R_{k^2}(h_\varepsilon) -R_{k^2}(h_0^D)||=0,
$$
\end{lemma}

\subsubsection{Preliminary lemma}

Let us introduce the Green functions associated to the resolvents  $R_{k^2}(h_0)$ and $R_{k^2}(h_0^D).$ There are the kernels
 $R(h_0)(x,y,k^2)$ and $R(h_0^D)(x,y,k^2)$ respectively.
To prove the lemma \ref{lemmeprincipal} we need the following lemma:
\begin{lemma}\label{lemma}
Let $\textbf{v}$ be a vector normalized to 1 and $P$ and $Q$ two projectors such that
\begin{equation}\label{hypo3}
P=(.,\textbf{v})\textbf{v},\quad Q=1-P,\quad \textbf{v}\in L^2(\mathbb{R}), \quad
\sup_{p\in \mathbb{R}}\widehat{V}(p)<\infty
\end{equation}
Let $\tau$ be the trace operator (and $\tau^\star$ its adjoint) acting as follow
$$
\tau f(x,y)=f(0,y)
$$
\noindent
Then\\
\noindent
(i)
$$
\lim_{\varepsilon \rightarrow 0}|| r_0  U_\varepsilon^\star \frac{v}{\sqrt{\varepsilon}}
 P \frac{v}{\sqrt{\varepsilon}}U_\varepsilon r_0-r_0\tau^\star \tau r_0||=0
$$
\noindent
(ii)
$$
\lim_{\varepsilon \rightarrow 0}||r_0 U_\varepsilon^\star v\frac{1}{\sqrt{\varepsilon}}
\frac{1}{\sqrt{\varepsilon}}v U_\varepsilon r_0-r_0\tau^\star\tau r_0||=0
$$
(iii)
$$
|| r_0 U_\varepsilon^\star \frac{v}{\sqrt{\varepsilon}} Q ||=o(\varepsilon)
$$
\end{lemma}
\noindent
\textbf{Proof}: to prove this lemma, we use the properties of the Fourier
transforms of the terms $r_0U^\star_\varepsilon \frac{v}{\sqrt{\varepsilon}}P
\frac{v}{\sqrt{\varepsilon}}U_\varepsilon r_0$, $r_0 U_\varepsilon^\star v
\frac{1}{\sqrt{\varepsilon}}\frac{1}{\sqrt{\varepsilon}}v U_\varepsilon r_0$
and $r_0\tau^\star\tau r_0$. Let us denote the Fourier transform $F$ and
its inverse $F^{-1}$ and recall
$$
(F\varphi)(p)=\widehat{\varphi}(p)=
\frac{1}{\sqrt{2\pi}}\int_\mathbb{R}e^{-ipx}\varphi(x)dx,\quad (F^{-1}\varphi)(x)=
\frac{1}{\sqrt{2\pi}}\int_\mathbb{R}e^{ipx}\varphi(p)dp
$$
\noindent
Let us do some useful calculations:
$$
(U_\varepsilon \varphi)(x)=
\frac{1}{\sqrt{\varepsilon}}\varphi(\frac{x}{\varepsilon})\nonumber\\
=\frac{1}{\sqrt{\varepsilon}}\int_\mathbb{R}\delta(\frac{x}{\varepsilon}-y)\varphi(y)dy
$$
The Fourier transform of a kernel $X$ is expressed as follow
$$
(FXF^{-1}\varphi)(p)=
\frac{1}{\sqrt{2\pi}}\int_\mathbb{R}dx
\,e^{-ipx}\int_\mathbb{R}dy\, X(x,y)\frac{1}{\sqrt{2\pi}}\int_\mathbb{R}e^{iqy}\varphi(q)dq.
$$
\noindent
Then, denoting $\widehat{U}_\varepsilon^\star (p,q)={\sqrt{2\pi}}\delta(\varepsilon q-p)$ we get
\begin{equation}
(FU^\star_\varepsilon F^{-1})(p)=
\int_\mathbb{R}\frac{\sqrt{\varepsilon}}{\sqrt{2\pi}}\delta(\varepsilon q-p)dq= :\int_\mathbb{R}\widehat{U}_\varepsilon^\star (p,q)dq.
\label{recap1}
\end{equation}
\noindent
Inserting the identity $FF^{-1}$ between the operators $U_\varepsilon^\star$ and $V$, we obtain
\begin{equation}\label{recap2}
FU_\varepsilon^\star VF^{-1}=\frac{\sqrt{\varepsilon}}{\sqrt{2\pi}}\widehat{V}(\varepsilon q)
\end{equation}
\noindent
Actually, we use the unitarity of the Fourier transforn $F$ and we insert the identity
$FF^{-1}$ on the terms listed above, we use \eqref{recap1} and \eqref{recap2}, and the fact that
$
\int_\mathbb{R}\widehat{V}(\varepsilon s)ds=1=\sqrt{2\pi}\widehat{V}(0).
$
Then, we get the following
unitary equivalences
\begin{equation}\label{proof of i}
||r_0U^\star_\varepsilon \frac{v}{\sqrt{\varepsilon}}P\frac{v}{\sqrt{\varepsilon}}
U_\varepsilon r_0||
= ||(.\widehat{r}_0,\frac{\widehat{V}(\varepsilon q)}{\sqrt{2\pi}})
\widehat{r}_0\frac{\widehat{V}(\varepsilon q)}{\sqrt{2\pi}}||, \quad
||r_0\tau^\star \tau r_0||= ||(.\widehat{r}_0,\widehat{V}_0)
\widehat{r}_0\widehat{V}_0||,
\end{equation}
\noindent
where we denote $\widehat{r}_0$ the Fourier transform of the resolvent $r_0.$
\noindent
and the fact that
$
\int_\mathbb{R}\widehat{V}(\varepsilon s)ds=1=\sqrt{2\pi}\widehat{V}(0).
$
\\
\noindent
\textbf{Proof} of (i).
\noindent
We only have to show, see \eqref{proof of i} that
$
\lim_{\varepsilon \rightarrow 0}||\frac{1}{\sqrt{2\pi}}\widehat{r}_0
\widehat{V}(\varepsilon q)-\widehat{r}_0\widehat{V}_0||=0.
$
Given that $\widehat{V}(\varepsilon q)$ converges pointwise to $\widehat{V_0}$. From the condition
\eqref{hypo3} and because the resolvent $r_0$ is compact we deduce that
$|\widehat{r}_0(q)(\widehat{V}(\varepsilon q)-\widehat{V_0})|$ is integrable in q.
Then $|\widehat{r}_0(q)(\widehat{V}(\varepsilon q)-\widehat{V}_0)|^2$ is bounded by an integrable
function in $q.$ The proof of (i) ended using the Lebesgues dominated convergence, that is to say,
\begin{equation}\label{V2}
\lim_{\varepsilon \rightarrow 0}\int_\mathbb{R}dq |\widehat{r}_0(q)(\widehat{ V}(\varepsilon q)
-\widehat{V_0})|^2=0.
\end{equation}
\noindent
\textbf{Proof} of (ii). First we rewrite
$
||r_0U_\varepsilon^\star v\frac{1}{\sqrt{\varepsilon}}\frac{1}{\sqrt{\varepsilon}}
vU_\varepsilon r_0||$ as $||\widehat{r}_0 FU_\varepsilon^\star \frac{V}{\varepsilon}
U_\varepsilon F^{-1}\widehat{r}_0||.$
Using the Fourier transform of
$
(FVF^{-1}\varphi)(p)$ given by $\frac{1}{\sqrt{2\pi}}\int_\mathbb{R}dq\, \varphi(q)\widehat{V}(p-q)$
and a straightforward computation we get
$
FU^\star_\varepsilon \frac{V}{\varepsilon}F^{-1}=\int_\mathbb{R}
\widehat{V}(\varepsilon (s-q) )dq,
$
\noindent
so that the kernel $\widehat{U_\varepsilon^\star V}(p,q)$ is
$
\varepsilon^{-1/2}\widehat{U_\varepsilon^\star V}(p,q)=
\widehat{V}(\varepsilon (p-q)).
$
\noindent
Then we have to prove the following convergence
$$
\lim_{\varepsilon \rightarrow 0}|\widehat{r_0}(p)
\left(\widehat{V}(\varepsilon (p-q))-\widehat{V}_0\right)\widehat{r_0}(p)|=0
$$
\noindent
$\widehat{V}(\varepsilon (p-q))$ converge point wise to $\widehat{V}_0$ and $|\widehat{r_0}(p)\left(\widehat{V}(\varepsilon (p-q))-
\widehat{V}_0\right)\widehat{r_0}(p)|$ is bounded an integrable function. As above, we use the Lebesgue dominated convergence
and we are done.
\\
\noindent
\textbf{Proof} of (iii). Let us use again the unitarity of the Fourier transform and equality \eqref{proof of i}.
We get the unitarity equivalence between $
||r_0U_\varepsilon^\star \frac{v}{\sqrt{\varepsilon}}
(1-P)\frac{v}{\sqrt{\varepsilon}}U_\varepsilon r_0||$ and
$
||\widehat{r}_0FU_\varepsilon^\star \frac{v}{\sqrt{\varepsilon}}(1-P)
\frac{v}{\sqrt{\varepsilon}}U_\varepsilon F^{-1}\widehat{r}_0||.$ We have to show that this term is $o(\varepsilon^2).$
\noindent
With the same tools, we compute:
$$
(\widehat{U_\varepsilon^\star V U_\varepsilon}\varphi)(p,q)
=\int_\mathbb{R}\widehat{V}(\varepsilon (p-q))\varphi (\varepsilon q)dq,
\,\,\,\, \mbox{and}\,\,\,\,(\widehat{\Pi_\varepsilon}\varphi)(p)
=\frac{1}{2\pi}
\int_\mathbb{R}\widehat{V}(\varepsilon p)\widehat{V}(-\varepsilon q)\varphi(q)dq
$$
\noindent
So, the kernel $(F \Pi_\varepsilon  F^{-1})(p,q)$ is given by $\widehat{V}(\varepsilon p)\widehat{V} (- \varepsilon q).$
\noindent
{From} the hypothesis on $V$ we knows that $xV(x)\in L^1(\mathbb{R})$.
We need to show
\begin{align}
&(a) \lim_{\varepsilon \rightarrow 0} |\frac{\widehat{V}(\varepsilon p)
\widehat{V} (- \varepsilon q)-\widehat{V_0}\widehat{V}(\varepsilon (p-q))}{\varepsilon}|
=0\quad \mbox{ almost everywhere}& \nonumber\\
&(b)\widehat{r_0}^2(p)(\frac{\widehat{\Pi}_\varepsilon(p,q)-
\widehat{V_0}\widehat{V}(\varepsilon (p-q))}{\varepsilon})|^2\widehat{r_0}^2(q)\,\,
\mbox{bounded by an integrable function in p and q} &\nonumber
\end{align}
\noindent
To check the point (a) we apply the mean value theorem, that is to say, since
$\widehat{V} (\varepsilon p)=\widehat{V}( 0)+\varepsilon p\,\, \widehat{V}'(\theta \varepsilon p),
\,\,  \forall \theta \in (0,1)$, then
\begin{eqnarray}
&&\widehat{V}(\varepsilon p)\widehat{V}(-\varepsilon q)-
\widehat{V_0}\widehat{V}(\varepsilon(p-q))=\nonumber\\
&&=(\widehat{V_0}+\varepsilon p\,\, \widehat{V}'(\theta \varepsilon p))
(\widehat{V_0}-\varepsilon q\,\, \widehat{V}'(\theta \varepsilon q))-
(\widehat{V_0}+\varepsilon (p-q)\,\, \widehat{V}'(\theta \varepsilon  (p-q)))
\widehat{V_0}\nonumber\\
&&=\varepsilon (p-q)\left(\widehat{V}'(\theta \varepsilon p)\widehat{V}'(\theta \varepsilon q)
-\widehat{V}'(\theta \varepsilon (p-q))\widehat{V_0}\right).\label{proceed2}
\end{eqnarray}
\noindent
Inserting this result \eqref{proceed2} in the limit $(a)$, then we are done.
\noindent
$V^{'}(p)$ is integrable in $p$ and $\widehat{r_0}^2(p)(p^\alpha q^\beta)\widehat{r_0}^2(q)$
for $0\leq \alpha, \beta\leq 2$ is integrable in $p$ and $q$ so the point $(b)$ is satisfied.\quad \QED \\
\noindent

\subsubsection{Proof of the lemma \ref{lemmeprincipal}}

\noindent
\textbf{Proof}. Recall the Green functions associated to the resolvents $R_{k^2}(h_0)$ and $R_{k^2}(h_0^D)$ as the
kernels $R(h_0)(x,y,k^2)$ and $R(h_0^D)(x,y,k^2)$ respectively. Using the resolvent equation, $R(h_0^D)(x,y,k^2)$ is computed as follow:
$$
R(h_0^D)(x,y,k^2)=R(h_0)(x,y,k^2)-C_k R(h_0)(x,0,k^2)R(h_0)(0,y,k^2), \quad C_k:=1/r_0(0,0,k^2).
$$
\noindent
The Green function
$r_0(x,y,k^2)$ expresses as
\begin{equation}\label{greenfunction1}
R(h_0)(x,y,k^2)=\sum_n \lambda_n^{-1}\psi_n(x)\psi_n(y),\quad \lambda_n=\alpha(n+\frac{1}{2}),
\end{equation}
\noindent
and denoting $H_n(x)$ the n-th Hermite polynomials,
\begin{equation}\label{greenfunction2}
\psi_n(x)=\sqrt{\frac{1}{\sqrt{\pi}2^n n!}}e^{-x^2/32}H_n(x).
\end{equation}
 Thanks to the symmetrized resolvent equation, we compute $R(h_\varepsilon)$ as
\begin{equation}
R(h_\varepsilon) =
R(h_0)-\frac{1}{\varepsilon^2}R(h_0)U_\varepsilon^\star \sqrt{V}
 T(\varepsilon k) \sqrt{V}U_\varepsilon R(h_0)\label{rvarepsilon},
\end{equation}
\noindent
where we denote $T(\varepsilon k)$ the following kernel
\begin{equation}\label{T(varepsilon k)}
T(\varepsilon k)=
\left(1+\frac{1}{\varepsilon^2} \sqrt{V}U_\varepsilon
R(h_0) (k) U_\varepsilon^\star \sqrt{V}\right)^{-1}.
\end{equation}
\noindent
We note that by a change of variable, we get the equality
\begin{equation}
\varepsilon^{-2}U_\varepsilon R_{k^2}(h_0)U_\varepsilon^\star f=
\int_{\mathbb{R}^2} R(h_0)(\varepsilon x,\varepsilon y,k^2)f(y)dx\,\,dy.
\end{equation}
\noindent
First, we show that we can decompose the kernel \eqref{T(varepsilon k)} as the sum of two terms, $t_0$ and $\varepsilon
t_1$ defined below, plus $t_{(2)}$, which are terms of order greater than or equal to $\varepsilon^2$. The most important 
part of the proof lies in the
fact that the Fourier transforms of $\varepsilon^{-1/2}R(h_0)U_\varepsilon^\star t_i^{1/2}, i=0,1, (2)$ is $o(\varepsilon)$
so that $t_1$ and $t_{(2)}$ does not contribute in the limit $\varepsilon$ goes to zero. Actually, formally we get
$$
\int_{\mathbb{R}^2}\frac{1}{\varepsilon}R(h_0)(x,y,k^2)U_\varepsilon^\star f(y,z)dydz=\int_{\mathbb{R}^2}R(h_0)
(\varepsilon x,\varepsilon y,k^2)f(y,z)dydz
$$
\noindent
which goes to $\int_{\mathbb{R}^2}R(h_0)(x,0,k^2)f(y,z)dydz$ as $\varepsilon$ goes to zero, and $\sqrt{V}t_0^{1/2}$ goes 
to a constant. So first, let us deal with $T(\varepsilon k)$ and show that it is invertible., More precisely we rewrite the kernel
$1/\varepsilon^2 U_\varepsilon^\star R(h_0) (x,y,k^2)$ using equation:
$$
R(h_0)(x,y;k^2)=R(h_0)(0,0,k^2)+\vec{x}.\vec{\nabla}R(h_0)(0,0,k^2)+
\vec{x}.\nabla^2 R(h_0)(0,0,k^2).\vec{x}+O(|x|^3).
$$
\noindent
Thanks to the definition of the green function, see for
example \cite{Kato}, we compute
$$
\vec{\nabla}R(h_0)(x,y,k^2)=\left\{\begin{array}{clrr}
 -\partial_xR(h_0)(x,y, k^2)+\partial_yR(h_0)(x,-y,k^2)\quad\mbox{if } \quad y\leq x \\
  -\partial_xR(h_0)(-x,y,k^2)+\partial_yR(h_0)(x,y,k^2) \quad\mbox{if } \quad y>x \\
\end{array}
\right.
$$
\noindent
So we get
$$
\vec{x}.\vec{\nabla}R(h_0)(0,0,k^2)= \left(\partial_yR(h_0)(0,0,k^2)+
\partial_xR(h_0)(0,0, k^2)\right)|x-y|.
$$
\noindent
This term does not have any singularity for $k^2$ close to zero thanks to the properties of $\eqref{greenfunction2}$.
Since we get $R(h_0)(x,y,k^2)=a+b|x-y|+O(|x^2|),\quad a,b\in \mathbb{R}.$
\noindent
then
$$
\sqrt{V}R(h_0)(\varepsilon x,\varepsilon y,k^2)\sqrt{V}=cP+\varepsilon M_1(x,y)+
M_{(2)}(x,y),
$$
\noindent
where
$$
P:=\frac{(., \sqrt{V})\sqrt{V}}{||V||}, \quad c:=a||V||, \quad M_1(x,y)= b\sqrt{V}|x-y|\sqrt{V},
$$
\noindent
and $M_{(2)}(x,y,k):=\sqrt{V}R(h_0)(\varepsilon x,\varepsilon y,k^2)\sqrt{V}-cP-M_1
=O(|\varepsilon^2x^2|).$ We also note that $\varepsilon M_1(x,y)=M_1(\varepsilon x, \varepsilon y)$.\\
\noindent
Using the Taylor Young formula, and the expression of the green function see \eqref{greenfunction1} and \eqref{greenfunction2} we get
$$
M_1(\varepsilon x, \varepsilon y)=\sqrt{V}R(h_0)(\varepsilon x, \varepsilon y,k^2)\sqrt{V}-cP
=o(1).
$$
The term $\frac{1}{\varepsilon^2} \sqrt{V}U_\varepsilon R(h_0) (k) U_\varepsilon^\star \sqrt{V}$
in \eqref{T(varepsilon k)} is $O(1)$ in $\varepsilon$ and so is $T(\varepsilon k).$ Indeed,
\begin{eqnarray}
(1+\sqrt{V}R(h_0)(\varepsilon x,\varepsilon y,k^2)\sqrt{V})^{-1}
&=&(1+cP)^{-1}\left(1-\varepsilon M_1(1+cP)^{-1}-M_{(2)}(1+cP)^{-1}\right)\nonumber\
\end{eqnarray}
\noindent
Rewriting $(1+cP)^{-1}$ as the sum $\sum_{k=0}^{\infty}(-cP)^k$, a straightforward calculation gives
$
(1+cP)^{-1}=
Q+c^{-1}P.
$
\noindent
Then we get the decomposition of $T(\varepsilon k)$ as the sum
\begin{equation}\label{T}
T(\varepsilon k)=t_0+\varepsilon t_1+O(\varepsilon^2),\quad t_0=
Q+\frac{1}{c}P,\quad t_1=(Q+\frac{1}{c}P)M_1(Q+\frac{1}{c}P).
\end{equation}
\noindent
The next step consists in showing the two following convergences as $\varepsilon$ goes to zero
\begin{eqnarray}
\frac{1}{\varepsilon^2}R(h_0)U_\varepsilon^\star \sqrt{V}t_0\sqrt{V}U_\varepsilon R(h_0)
&\rightarrow& CR(h_0)(x,0,k^2)R(h_0)(0,y,k^2)\nonumber\\ \frac{1}{\varepsilon}R(h_0)
U_\varepsilon^2 \sqrt{V} t_1\sqrt{V}U_\varepsilon R(h_0)
&\rightarrow& 0.
\end{eqnarray}
\noindent
\noindent
Going back to \eqref{rvarepsilon} and \eqref{T} we get
\begin{eqnarray}
R(H_\varepsilon)&=&R(h_0)-\frac{1}{\varepsilon^2}R(h_0)U_\varepsilon^\star \sqrt{V} \left(t_0(k)+
\varepsilon t_1(k)+O(\varepsilon^2 k^2)\right) \sqrt{V}U_\varepsilon R(h_0)\nonumber\\
&=&R(h_0)-\frac{1}{\varepsilon^2}R(h_0)U_\varepsilon^\star \sqrt{V} \left(Q+\frac{P}{c}+
\varepsilon (Q+\frac{P}{c})M_1(Q+\frac{P}{c})+O(\varepsilon^2 k^2)\right)
\sqrt{V}U_\varepsilon R(h_0)\nonumber
\end{eqnarray}
\noindent
Then,
$$
\lim_{\varepsilon \rightarrow 0}||R(H_\varepsilon)-R(h_0)||=
\lim_{\varepsilon \rightarrow 0}||\frac{1}{\varepsilon^2}R(h_0)U_\varepsilon^\star \sqrt{V}
\frac{P}{c} \sqrt{V}U_\varepsilon R(h_0)||\frac{1}{R(h_0)(0,0,k^2)}R(h_0)\tau^\star \tau R(h_0).
$$
\noindent
Using the point (iii) of the lemma \ref{lemma} and the fact that $M_1$ is bounded
we get that
$||\varepsilon^{-2} R(h_0)U_\varepsilon^\star \sqrt{V} Q\sqrt{V}U_\varepsilon R(h_0)||,
 ||\varepsilon^{-1} R(h_0)U_\varepsilon^\star \sqrt{V}QM_1Q\sqrt{V}U_\varepsilon R(h_0)||$ and\\
 \noindent
 $||\varepsilon^{-1} R(h_0)U_\varepsilon^\star \sqrt{V}QM_1\frac{P}{c}\sqrt{V}U_\varepsilon R(h_0)||$
 go to zero as $\varepsilon$ goes to zero.
 From the point (i) we show
 $||R(h_0)-\varepsilon^{-2}R(h_0)U_\varepsilon^\star \sqrt{V} P/c\sqrt{V}U_\varepsilon R(h_0)||$
 goes to zero and we are done.\quad \QED

\section{Concluding remarks}

\begin{figure}[h]\label{figure 2}
\begin{center}
\includegraphics[scale=0.21]{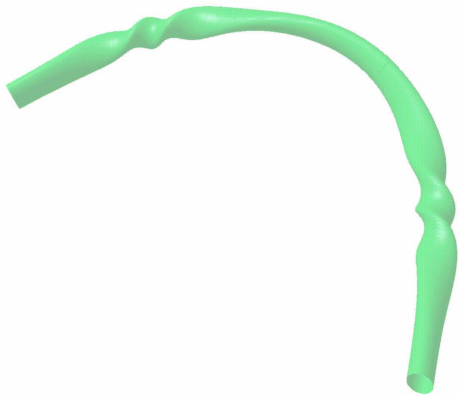}
   \vspace{1cm}
   \caption{An example of twisted and bent waveguide}
\end{center}
\end{figure}

In this paper we addressed to the question of operator-norm resolvent convergence of the one-particle Hamiltonian
in the limit of shrinking wave-guide and scaled twisting.

The question of the validity of the norm-resolvent convergence and the idea of this paper are due to
Pierre Duclos and David Krej\v{c}i\v{r}\'{\i}k. This problem was explicitly raised in \cite{K-talk} and then treated in 
the context of thin quantum wave-guides in \cite{KS},\cite{KrSe}, under regularity conditions different then ours.

The three-step strategy of the proof we proposed in Section 3 gives the $O(\varepsilon)$ rate for convergence to 
the limiting operator. Apparently this is not an optimal estimate. Therefore, one of the open question is relaxing the
conditions of our main Theorem versus optimality of the rate. Another aspect is to compare our strategy and conditions
with those of \cite{KS},\cite{KrSe}.

Twisting versus bending in the limit of thin quantum wave-guides, see for example Fig.2, is an open question that 
definitely merits special attention. A progress in this direction due to the Hardy inequality technique \cite{Kr08} 
is apparently a good basis to study this problem.

\section{Acknowledgments}

This paper is dedicated to my teacher and supervisor
Professor Pierre Duclos (1948-2010). He was fascinated by this subject during his last years. With a delicate persistence he was 
trying to teach me to share with him this fascination, the beauty of arguments and motivations, rooted in his passion for mathematical physics.
With readiness and stubbornness he spent time understanding and answering my questions during our long discussions.

I would like to express my profound gratitude to David Krej\v{c}i\v{r}\'{\i}k. His discussions with Pierre, and then with me in Prague and
Marseille, his advices on different stages of preparation of the manuscript for publication were extremely helpful and indispensable to 
make this project possible.

I am also grateful to H.\v{S}ediv\'{a}kov\'{a} for useful discussions on the subject and results of her diploma thesis \cite{KrSe}.

Finally, I thank Valentin Zagrebnov for our discussions, for his advices and help with this project as well as for his support 
during these last two years.


\begin{thebibliography}{999}
\bibitem{ACF}
S.Albeverio, C. Cacciapuoti, D.Finco,  \textit{Coupling on the singular limit of thin quantum waveguides}, 
J. Math. Phys. \textbf{48}, 032103, (2007).

\bibitem{BGW}
D.Boll\'{e}, F.Geszetesy, S.F.J Wilk, \textit{A complete treatment of low-energy scattering in one dimension}, J.Oper.Theory \textbf{13}
(1985),3 -31.

\bibitem{CDFK}
B. Chenaud, P.Duclos, P. Freitas, D. Krej\v{c}i\v{r}\'{\i}k , \textit{Geometrically induced discrete spectrum in curved tubes}, 
Differential Geom. Appl. \textbf{23}, no. 2, pp. 95-105, (2005).

\bibitem{CE}
C.Cacciapuoti, P.Exner, \textit{Non trivial edge couling from a Dirichlet network squeezing: the case of a bent waveguide},
J.Phys. A: Math. Theor. \textbf{40} (2007) F511-F523.

\bibitem{DuEx}
P.Duclos, P.Exner, \textit{Curvature-induced bound states in quantum waveguides in two and three dimensions}, 
Reviews in Mathematical Physics, 7:73, 102, (1995).

\bibitem{EKK}
T.~Ekholm, H.~Kova{\v{r}}{\'\i}k, and D.~Krej\v{c}i\v{r}\'{\i}k, \emph{A
  {H}ardy inequality in twisted waveguides}, Arch. Ration. Mech. Anal.
  \textbf{188} (2008), 245--264.

\bibitem{ExSh}
P.Exner, Seba, \textit{Bound states in a curved wave guide}, J.Math. Phys. \textbf{30}, 2574-2580, (1989).

\bibitem{FS}
L.Friedlander, M.Solomyak , \textit{On the spectrum of the Dirichlet Laplacian in a narrow strip},
Israeli Math. J.170 (2009), no 1,337-354.

\bibitem{JN}
A.Jensen, G.Nenciu,\textit{ A unified approach to resolvent expansions at threshold}
Rev. Math. Phys. 13, 717-754 (2001).

\bibitem{Kato}
T. Kato, \textit{Perturbation Theory for Linear Operators, Springer-Verlag Berlin}, Heidelberg, New York, (1966).

\bibitem{KK3}
M. Kolb, D.~Krej\v{c}i\v{r}\'{\i}k,
\emph{The Brownian traveller on manifolds},
preprint on arXiv:1108.3191 [math.AP] (2011).


\bibitem{KK2}
H.~Kova\v{r}{\'\i}k and D.~Krej\v{c}i\v{r}\'{\i}k, \emph{{A Hardy inequality in
  a twisted Dirichlet-Neumann waveguide}}, Math. Nachr. \textbf{281} (2008),
  1159--1168.

\bibitem{K3}
D.~Krej\v{c}i\v{r}\'{\i}k, \emph{Hardy inequalities in strips on ruled
  surfaces}, J. Inequal. Appl. \textbf{2006} (2006), Article ID 46409, 10
  pages.

\bibitem{Kr08}
D. Krej\v{c}i\v{r}\'{\i}k, \textit{Twisting versus bending in quantum waveguides, Analysis on Graphs and its
Applications}, Proceedings of Symposia in Pure Mathematics, American Mathematical Society \textbf{15} (2008)
555--568. See arXiv:0712.3371v2[math-ph] for corrected version.

\bibitem{K-talk}
D. Krej\v{c}i\v{r}\'{\i}k,
\emph{The Hardy inequality and the asymptotic behaviour of
the heat equation in twisted waveguides},
talk at the conference in honor of Pierre Duclos,
Quantum Dynamics, Marseille, November 2010.

\bibitem{KS}
D. Krej\v{c}i\v{r}\'{\i}k, H. \v{S}ediv\'{a}kov\'{a},
\emph{The effective Hamiltonian in curved quantum waveguides
under mild regularity assumptions}, preprint.

\bibitem{KrZu10}
D. Krej\v{c}i\v{r}\'{\i}k, E.Zuazua, \textit{The Hardy inequality and the heat equation in twisted tubes}, J.Math.Pure Appl. 
\textbf{94} (2010) 277-303.
\bibitem{KrZu11}
D. Krej\v{c}i\v{r}\'{\i}k, E.Zuazua, \textit{The asymptotic behabiour of the heat equation in a Dirichlet-Neumann waveguide}, 
J. Differential Equations 250, 2334-2346, (2011).

\bibitem{K}
P.Kuchment, \textit{Graphs models for waves in thin structures, published in Waves in Random media}, 12(2002), no. 4,R1-R24.


\bibitem{KrSe}
H.\v{S}ediv\'{a}kov\'{a}, \textit{Quantum Waveguides under Mild Regularity Assumptions}, Diploma Thesis,
Czech Technical University - Prague, FJFI (2011).

\end{thebibliography}
\end{document}